\documentclass[useAMS,usenatbib]{mn2e}
\usepackage{graphics,epsfig,amsmath,amssymb,amstext,txfonts,array,color}
\usepackage{graphicx}
\usepackage[T1]{fontenc}

\usepackage{color}

\newcommand\simlt{\lower.5ex\hbox{$\; \buildrel < \over \sim \;$}}
\newcommand\simgt{\lower.5ex\hbox{$\; \buildrel > \over \sim \;$}}

\title[Dynamics of dissipative Poynting dominated flows]
{The effect of Compton drag on the dynamics of dissipative Poynting dominated 
flows:  Implications for the unification of radio loud AGN}

\author[Amir Levinson, Noemie Globus]
  {A.~Levinson,$^1$\thanks{E-mail: Levinson@wise.tau.ac.il}
  N.~Globus,$^2$  \\
  $^1$Raymond and Beverly Sackler School of Physics \& Astronomy, Tel Aviv University, Tel Aviv 69978, Israel\\
  $^2$Racah Institute of Physics, The Hebrew University of Jerusalem, 91904 Jerusalem, Israel}
\date{Released \today}

\pagerange{\pageref{firstpage}--\pageref{lastpage}} \pubyear{2002}

\def\LaTeX{L\kern-.36em\raise.3ex\hbox{a}\kern-.15em
    T\kern-.1667em\lower.7ex\hbox{E}\kern-.125emX}

\begin{document}
\label{firstpage}
\maketitle

\begin{abstract}
  The dynamics of a dissipative Poynting dominated flow subject to a radiation drag due 
to Compton scattering of ambient photons by relativistic electrons accelerated in reconnecting current 
sheets is studied.   It is found that the efficiency at which magnetic energy is converted to radiation is 
limited to a maximum value of $\epsilon_c=3l_{dis}\,\sigma_0/4(\sigma_0+1)$, where $\sigma_0$ 
is the initial magnetization of the flow and $l_{dis}\le1$ the fraction of initial Poynting flux that can dissipate.
The asymptotic Lorentz factor satisfies $\Gamma_\infty\ge\Gamma_0(1+l_{dis}\,\sigma_0/4)$, where $\Gamma_0$
is the initial Lorentz factor.   This limit is approached in cases where the cooling time is shorter than the local 
dissipation time.   A somewhat smaller radiative efficiency is expected 
if radiative losses are dominated by synchrotron and SSC emissions.   It is suggested that under certain conditions 
magnetic field dissipation may occur in two distinct phases: 
On small scales, asymmetric magnetic fields that are advected into the polar region and dragged out by the outflow
dissipate to a more stable configuration.  The dissipated energy is released predominantly as gamma rays. 
On much larger scales, the outflow encounters a flat density profile medium and re-collimates.  This leads
to further dissipation and wobbling of the jet head by the kink instability, 
as found recently in 3D MHD simulations.  Within the framework of a model proposed recently to explain the 
dichotomy of radio loud AGN, this scenario can account for the unification of  gamma-ray blazars with 
FRI and FRII radio sources. 
\end{abstract}
\begin{keywords}.
galaxies: active - quasars: general - radiation mechanism: nonthermal - gamma-rays: galaxies - galaxies: jets
\end{keywords}

\begin{figure*}
\centerline{\includegraphics[scale=0.36]{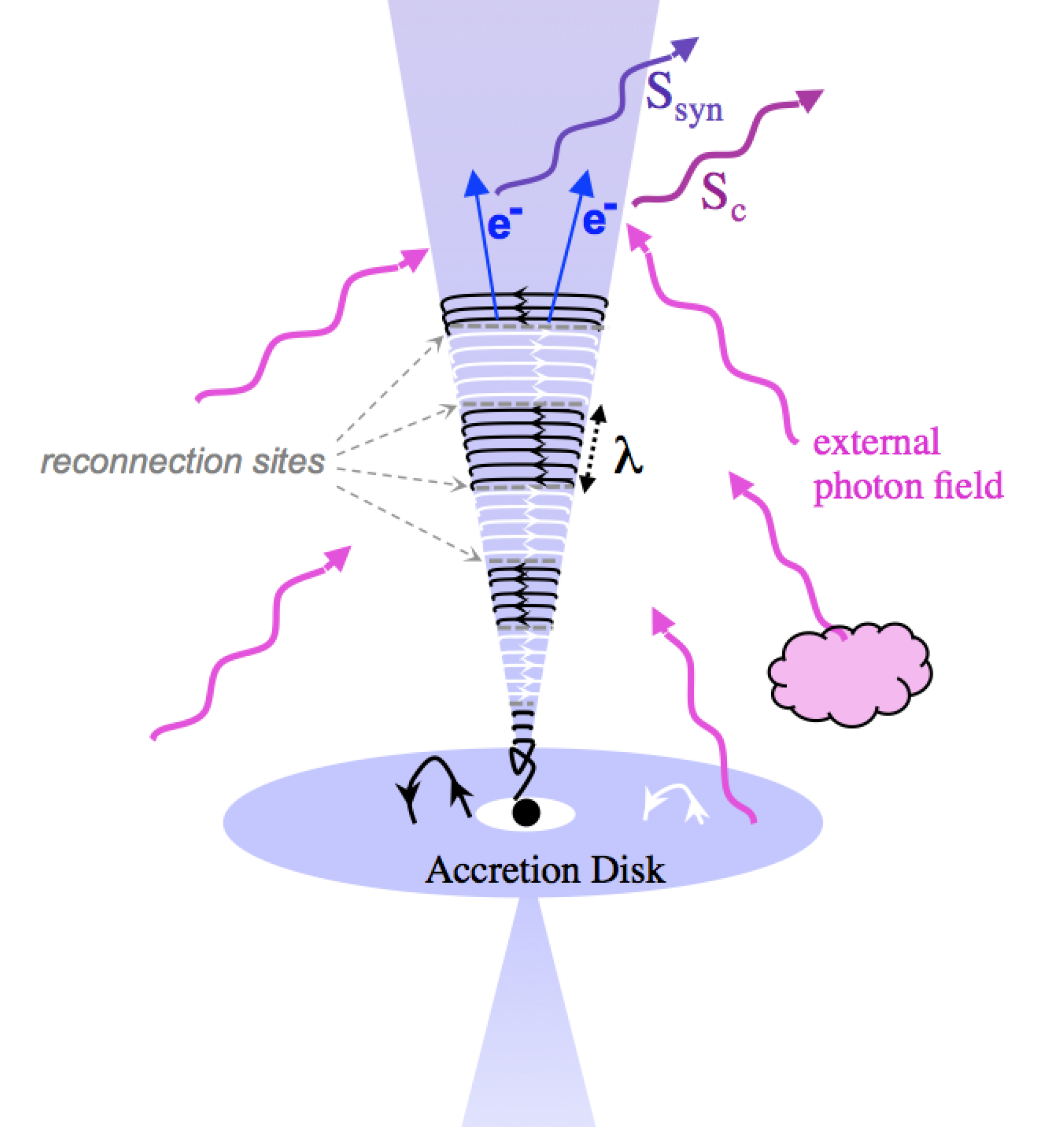}}
\caption{Schematic illustration of the outflow model:   Advection of asymmetric magnetic field into the black hole gives rise to 
formation of magnetic domains in the outflow across which the magnetic field changes polarity (indicated by the black and white 
stripes).  Collisionless reconnection in the current sheets separating those domains leads to acceleration of electrons to nonthermal energies.
Rapid cooling of the accelerated electrons, via synchrotron emission and inverse Compton scattering of external seed photons, gives rise 
to effective conversion of the dissipated energy into gamma radiation.}
\label{fig-f0}
\end{figure*}

\section{Introduction}

A key issue in the theory of magnetized outflows is the dissipation of magnetic energy.  A plausible
dissipation mechanism commonly invoked
is magnetic reconnection (Romanova \& Lovelace 1992, Levinson \& van Putten 1997, Drenkhahn \& Spruit 2002 (DS02); Lyutikov \& Blandford 2003; 
Giannios \& Spruit 2007; Lyubarsky 2010; McKinney \& Uzdensky 2012, Levinson \& Begelman 2013, 
Bromberg \& Tchekhovskoy 2015).   This mechanism requires
formation of small scale magnetic domains with oppositely oriented magnetic field lines.   Such 
structures may inherently form during outflow injection, as in the striped wind model, or result from 
current-driven instabilities induced during the propagation of the jet (Mignone et al. 2010, Mizuno et al. 2012, O'Neill et al. 2012, 
Guan et al. 2014, Bromberg \& Tchekhovskoy 2015). 

DS02 constructed a model for a dissipative Poynting-flux dominated outflow in GRBs, assuming that magnetic reconnection 
proceeds at a rate governed by the Alfv\'en speed, and allowing for isotropic emission in the outflow rest frame.   
They have shown that the release of magnetic energy gives rise to effective acceleration of the jet even in 
the presence of strong radiative losses.  In the optimal case, the efficiency at which Poynting energy is converted
to nonthermal radiation can 
approach 50\%.  The remaining Poynting energy is converted to bulk kinetic energy.  
This upper limit on the radiative efficiency is a consequence of the kinematic conditions.   
It can be achieved provided that nearly all the Poynting energy dissipates above the photosphere, on scales
at which the cooling rate exceeds the local dissipation rate.  

In certain circumstances the flow may be subject to a strong radiation drag (Phinney 1987, Li et al. 1992, 
Sikora et al. 1996, Beskin et al. 2004, Levinson 2007, Golan \& Levinson 2015).   This may be the case, e.g.,  
in powerful AGN and in microquasars.  The loss of bulk momentum by radiative friction should lead to reduced 
acceleration of the Poynting dominated flow and a higher radiative efficiency that can exceed the limit found in DS02.
In this paper we extend the model outlined in DS02 to flows propagating in an ambient radiation field, by incorporating 
source terms that account for scattering of external photons by electrons accelerated in reconnecting current sheets.  
We solve the dynamical equations numerically and compare the numerical solutions with an analytic solution obtained 
in the limit of rapid cooling. We also derive analytic expressions for the maximum radiative efficiency, and 
the asymptotic bulk Lorentz factor and magnetization of the outflow.

It has been proposed recently (Tchekhovskoy \& Bromberg, 2015) that re-collimation of Poynting dominated jets by ambient gas 
on kpc scales may explain the FRI- FRII dichotomy of radio loud AGN.  In this model, 
objects having a moderate jet power are significantly slowed down by the ambient medium and 
appear as FRI sources, owing to a rapid growth of the kink instability, whereas objetcs having powerful jets
are less susceptible to the instability and, therefore, 
keep propagating at relativistic speeds, forming strong shocks and backflows near the jet head, as
seen in FRII sources.  However, this scenario ignores the fact that a considerable fraction 
of the bulk energy must dissipate already on much smaller scales.  According to the unified model of 
radio loud AGN, FRI and FRII sources are associated with blazars when observed at small viewig angles
to the jet axis.  The strong, highly variable gamma-ray emission observed in many blazars seem to imply
high conversion efficiency of Poynting flux to gamma-ray emission on subparsec and parsec scales.
If dissipation on those scales is due to internal kink instability, e.g., owing to collimation by disk winds, 
then one naively expects that by the time the jet reaches kpc scales it will become weakly magnetized, unless
fine tuning of external conditions is invoked.  Below we propose 
that dissipation of magnetic energy might naturally occur in two stages, if an unstable magnetic field
configuration is established during the injection of the jet.  This is expected in cases where the magnetic field 
advected inwards by the accretion flow has substantial asymmetries.  The unstable magnetic field 
configuration in the jet
would then tend to relax to a more stable configuration over scales of the order of the characteristic size
of striped layers, typically hundreds of gravitational radii, thereby giving rise to the beamed
emission observed in blazars.  Our analysis 
indicates high radiative efficieny on these scales.  It also shows that after relaxing to its
stable state, the jet can remain magnetically dominated.  Thus, when encountering the confining medium on kpc scales,
it can follow the evolution predicted in Tchekhovskoy \& Bromberg (2015).

\section{The model}

We adopt  the  wind model of DS02, in which magnetic energy is dissipated locally through reconnection during the propagation  of 
the flow (figure \ref{fig-f0}).  Reconnection commences  at some radius $r_0$ at which the 4-velocity of the flow equals $u_0$,  
following an initial acceleration phase in the ideal MHD limit.    Local dissipation occurs over a  
time scale $\tau=(\lambda/c) \Gamma^2\epsilon^{-1}\sqrt{1+u_A^{-2}}$,
where $\lambda$ is the characteristic size of the reconnection layer (that is, the distance 
between neighboring stripes of different magnetic field orientation), $\Gamma=\sqrt{1+u^2}$ is the bulk 
Lorentz factor of the flow, $u_A$ is the local Alfv\'en 4-velocity in the comoving frame, and $\epsilon<1$ 
is the ratio of the reconnection and Alfv\'en speeds, with $\epsilon\simeq 0.1$ indicated by 
recent numerical simulations of relativistic reconnection.   The scale $\lambda$ depends on the magnetic
structure of the accretion flow, and is poorly constrained.  For illustration we adopt $\lambda=M$, where $M$ is the geometric
mass of the black hole.  As explained in DS02, the fraction of the initial Poynting flux that can dissipate depends on the magnetic field
configuration.     Various estimates of the magnetization in the emission zones of blazars, 
as well as energy considerations, seem to suggest that this fraction must be substantial.   Here we 
suppose that dissipation ceases sharply when the magnetization drops below some critical 
value $\sigma_c$.  The evolution of the magnetic field is then governed by the equation

\begin{equation}
\partial_r\ln(rbu)=-\frac{\theta(1-\sigma_c/\sigma)}{c\tau\beta}=-\frac{1}{\beta  \delta_B}\left(\frac{\Gamma_0}{\Gamma}\right)^{2}
\frac{\theta(1-\sigma_c/\sigma)}{\sqrt{1+u_A^{-2}}}, 
\end{equation}
where $\theta(z)$ is a step function, and we define the fiducial length scale 
\begin{equation}
\delta_B= \lambda\Gamma_0^2\epsilon^{-1}\simgt10^{17}M_9(\epsilon/0.1)^{-1}(\Gamma_0/10)^2\quad {\rm cm}\,.
\end{equation}
In regions of high magnetization $u_A>>1$, and we shall henceforth approximate $\sqrt{1+u_A^{-2}}=1$.
With these approximations, the equations governing the dissipative flow are:
\begin{eqnarray}
& &\frac{1}{r^2} \partial_r[r^2(w^\prime+b^2)\Gamma u]=S^0,\label{L}\\
& &\frac{1}{r^2} \partial_r[r^2(w^\prime+b^2)u^2+r^2b^2/2]+\partial_r p^\prime=S^r,\\
& & \partial_r\ln(rbu)=-\frac{1}{\beta  \delta_B}\left(\frac{\Gamma_0}{\Gamma}\right)^{2}\theta(1-\sigma_c/\sigma),\label{Bdiss}\\
& & \partial_r(r^2\rho^\prime u)=0.\label{dotM}
\end{eqnarray}
Here $w^\prime=\rho^\prime+p^\prime+e^\prime$ is the proper specific enthalpy, $p^\prime$ the pressure, 
$\rho^\prime$ the proper density, $e^\prime$ the internal energy density, $b=B^\prime/\sqrt{4\pi}$ the normalized, proper magnetic 
field, $\Gamma$ is the bulk Lorentz factor and $u=\Gamma\beta$ the 4-velocity.  The source terms $S^0$ and $S^r$ account
for energy and momentum losses, respectively.   
In what follows we adopt a relativistic equation of state, whereby $e^\prime=3p^\prime$.  
The integral of Equation (\ref{dotM}) gives the conserved mass flux,
\begin{equation}
\dot{M}=r^2\rho^\prime u.
\end{equation}
The total outflow power and Poynting power are given, respectively, by 
\begin{eqnarray}
& & L_j(r)=r^2(w^\prime+b^2)\Gamma u,\\
& & L_b=r^2b^2\Gamma u.
\end{eqnarray}

A fraction of the magnetic energy dissipated in the reconnection layer is tapped for acceleration of electrons to nonthermal energies.
Numerical simulations of collisionless magnetic reconnection in electron-positron plasma (Cerutti et al. 2012, Sironi \& Spitkovsky 2014, Werner et al. 2014, Kagan et al. 2015) indicate a power law energy distribution, $dn^\prime_e/d\gamma=\kappa_e\gamma^{-q}$;  
$\gamma_{1}<\gamma<\gamma_{2}$, with index $q$ that depends on the magnetization parameter of the flow outside the 
current sheet, $\sigma=b^2/(\rho^\prime c^2)$.  For $\sigma\simgt10$ the spectral  index lies in the range  $1< q < 2$.
Electron acceleration to nonthermal energies is expected also in case of reconnection in electron-proton plasma (Melzani et al. 2014, Sironi et al. 2015).
Henceforth, we employ the parametrization  $\xi_e=u^\prime_e/e^\prime$, where $u'_e=\int m_ec^2\gamma dn'_e$ is the total 
energy density of the nonthermal population, as measured in the comoving frame.   If equipartition between electrons and protons 
is established, as seems to be indicated by recent PIC simulations of reconnection in 
electron-proton plasma  (Melzani et al. 2014, Sironi et al. 2015), then $\xi_e\simeq0.5$.   For a flat 
distribution, $q<2$, the maximum energy of accelerated electrons, $\gamma_2$, is limited by the energy budget.   
Specifically, for an electron-proton plasma in rough equipartition it is 
given by $\gamma_2\simeq (m_p/2m_e)\sigma$ if $q\simeq1$, and may be considerably higher for a steeper
distribution (Melzani et al. 2014), with $\gamma_2\sim (m_p/m_e)[(\sigma+1)(2-q)/(2q-2)]^{1/(2-q)}$ 
for $1< q\simlt 2$.  Thus, for $\sigma >$
a few, we anticipate $\gamma_2\simgt10^5$ in electron-proton plasma, consistent with observations of gamma-ray blazars.

Compton scattering of ambient photons by the nonthermal
electrons accelerated in reconnection sites imposes a drag force on the flow.  We denote the total energy density 
of the external radiation field by $u_s(r)=u_{s0}f_s(x)$, where $x=r/\delta_B$.   To order $O(\Gamma^{-2})$ the source 
terms associated with the Compton drag are given by (Golan \& Levinson 2015)
\begin{eqnarray}
& & S^0_c=-\frac{8}{3}\Gamma^3<\gamma^2>u_s\sigma_T n^\prime_{e},\\ \label{S0c}
& & S_c^r=\beta S_c^{0}+S_c^0/3\Gamma^2,
\end{eqnarray}
in terms of the total electron density 
\begin{equation}
n^\prime_e=\int_{\gamma_{1}}^{\gamma_{2}} {\frac{dn^\prime_e}{d\gamma} d\gamma},\quad
\end{equation}
and the second moment
\begin{equation}
<\gamma^2>=\frac{1}{n'_e}\int_{\gamma_1}^{\gamma_2} {\gamma^2 } {\frac{dn^\prime_e}{d\gamma} d\gamma}.
\end{equation}
The first and second moments of the electron energy distribution are related to the maximum 
energy via $<\gamma^2>/<\gamma>=\chi\gamma_2$, here 

\begin{equation}
\chi=\frac{(2-q)}{(3-q)}\frac{[1-(\gamma_{2}/\gamma_{1})^{3-q}]}{[1-(\gamma_{2}/\gamma_{1})^{2-q}]}\frac{\gamma_1}{\gamma_2}.
\label{chi}
\end{equation} 
For the flat spectra observed in simulations of relativistic reconnection we estimate $\chi\simeq (2-q)/(3-q)$.  
Using the above parametrization, the energy source term associated with inverse Compton emission can be written as 
\begin{eqnarray}
S^0_c=-\frac{\alpha f_s(x)(\Gamma/\Gamma_0)e}{ \delta_B},
\end{eqnarray}
in terms of the internal energy density $e=\Gamma^2 e^\prime$, and  the dimensionless parameter
\begin{eqnarray}
\alpha&=&\frac{8\delta_B\sigma_T\chi\xi_{e}\Gamma_0\gamma_{\rm 2}u_{s0}}{3m_ec^2}\nonumber\\
&=&
32\chi\xi_{e}\left(\frac{l_B}{10^{17}\ {\rm cm}} \right)\left(\frac{\gamma_2}{10^5} \right)\left(\frac{\Gamma_0}{10}\right)\left(\frac{u_{s0}}{10^{-3}\ {\rm erg}\ {\rm cm}^{-3}}\right).
\label{alpha}
\end{eqnarray}
Note that $\alpha$ is roughly the ratio of the dissipation time at the initial radius $r_0$, $\tau_0=\delta_B/c$, and the cooling time
$t_c=3m_ec/(4\Gamma_0\gamma_2\sigma_Tu_{s0})$. Specifically   $\alpha=2\chi\xi_e\tau_0/t_c$.  As explained above, in general
$\gamma_2$ depends on the local magnetization $\sigma$.   However, as will be shown below,  in the
regime $\alpha f_s(x)>1$ the solution is highly insensitive to the value of $\alpha$, and   
we shall henceforth assume, for simplicity, that $\gamma_2$ is constant.   
\begin{figure*}
\centerline{\includegraphics[scale=0.5]{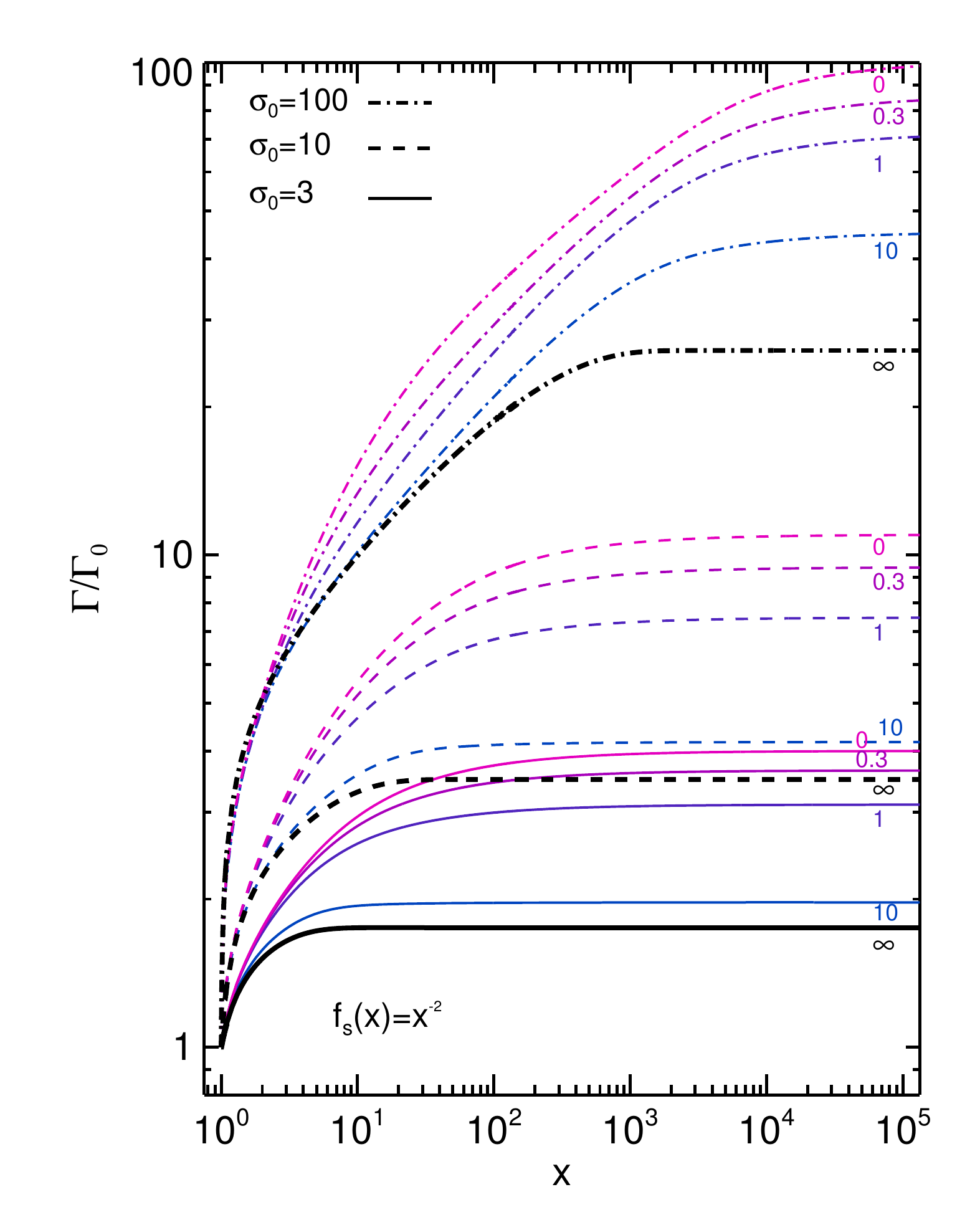}\includegraphics[scale=0.5]{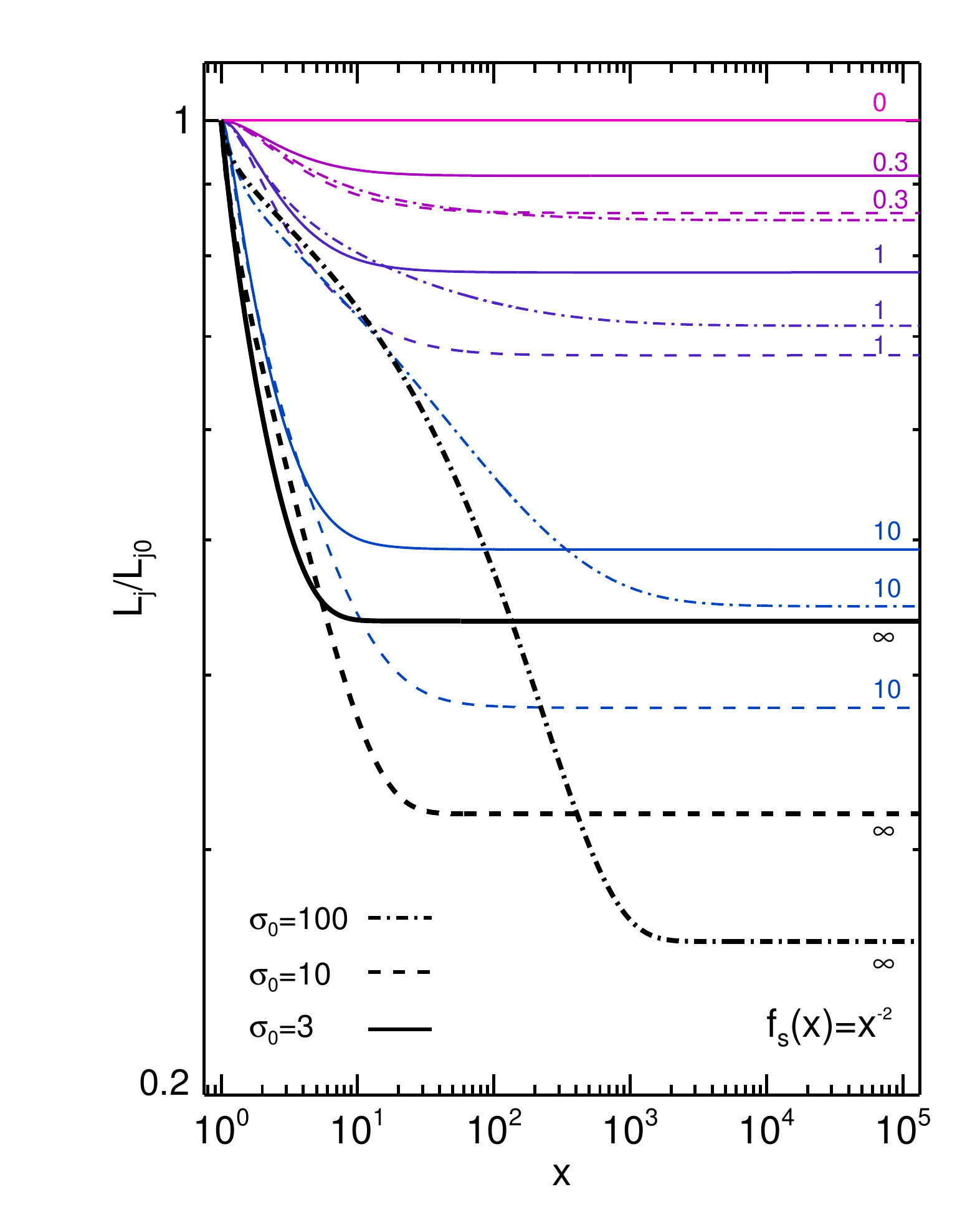}}
\caption{Profiles of the normalized Lorentz factor (left) and total jet power (right),  computed numerically using 
Equations (\ref{dPsi})-(\ref{deta}) for ambient radiation intensity profile $f_s(x)=x^{-2}$, and different values of 
$\alpha$ and the initial magnetization $\sigma_0$.  Different line types corespond 
to different values of $\sigma_0$, as indicated.  The numbers that label the curves are the values of $\alpha$.
The thick black lines delineate the analytic solution, Equations (\ref{Psi_analyt})-(\ref{L_B_analyt}). }
\label{fig-f1}
\end{figure*}

Synchrotron radiation also contributes to energy losses.  We denote by $S^0_{syn}$ the source term associated with synchrotron losses.
Since the emission is isotropic in the rest frame of the flow one readily obtains $S^r_{syn}=\beta S^0_{syn}$.   The total losses are 
given by  $S^0=S^0_c+S^0_{syn}$, $S^r=\beta S^0+S^0_c/3\Gamma^2=\beta S^0+\xi_cS^0/3\Gamma^2$, where 
we define $\xi_c=S^0_c/S^0$.

Equations (\ref{L})-(\ref{dotM}) can be rendered dimensionless upon using the normalization 
$l_j(r)=L_j(r)/L_{j0}$, $l_b(r)=L_b(r)/L_{b0}$, $\bar{\Gamma}(r)=\Gamma/\Gamma_0$,
where subscript $0$ denotes values at $r=r_0$, and defining the constant fractions $\kappa_B=L_{b0}/L_{j0}$,  
and $\kappa_w=\dot{M}c^2\Gamma_0/L_{j0}$.  In terms of the dimenssionless coordinate $x=r/\delta_B$ one then obtains
\begin{eqnarray}
\frac{dl_j}{dx}&=& -\frac{3}{4\xi_c}\alpha \bar{\Gamma} f_s(x)\left(l_j - \kappa_B l_b - \kappa_w \bar{\Gamma} \right),    \label{dPsi}\\
\frac{d\ln\bar{\Gamma}}{dx}&=& \frac{ \displaystyle(4\xi_c/3-1)\frac{dl_j}{dx}-\kappa_B \frac{dl_b}{dx} + \frac{2}{x}\left(l_j - \kappa_B l_b - \kappa_w \bar{\Gamma}\right) }{\kappa_w \bar{\Gamma}+ 2 l_j - 2 \kappa_B l_b}, \label{dGamma}\\
\frac{dl_b}{dx}&=& - \frac{2l_b}{\bar{\Gamma}^2}\theta(1-\sigma_c/\sigma), \label{deta}
\end{eqnarray}
subject to the initial conditions $l_j(x_0)=l_b(x_0)=\tilde{\Gamma}(x_0)=1$.  In deriving Equation (\ref{deta}) 
we invoked the approximation $L_b=r^2b^2u^2$ that holds in the relativistic limit.

\section{Results}

We integrate Equations (\ref{dPsi})-(\ref{deta}), starting at $x=1$ ($r=r_0=\delta_B$), where $\Gamma=\Gamma_0$ and $l_j=l_b=1$.   
The plasma is assumed to be cold at the injection point, in which case $\kappa_B=1-\kappa_w=\sigma_0/(1+\sigma_0)$, 
where  $\sigma_0=\kappa_B/\kappa_w$ is the initial magnetization.  We consider first the limit where energy losses are
dominated by Compton scattering of external radiation, and set $\xi_c=1$. 
We examin two models for the intensity profile of the external radiation which is intercepted by the flow.
In the first one it scales like that of a point source, $f_s(x)=x^{-2}$.  In the second one
$f_s(x)=1$ at $1<x<x_e$ and $f(x)=(x/x_e)^{-2}$ at $x>x_e$.  The latter choice 
is motivated by detailed calculations of the seed photon field contributed by extended radiation sources in blazars (Joshi et al. 2014).

When $\alpha>>1$ the cooling time is much shorter than the dissipation time of magnetic energy.  Then, the 
dissipated energy is radiated away instantaneously, keeping the internal energy small, $e^\prime<<\rho^\prime$, so
that $L_j\simeq L_b+\dot{M}c^2\Gamma$ .   To order $O(\alpha^{-1})$
Equations (\ref{dPsi}) and (\ref{dGamma}) admit  the analytic solution
\begin{eqnarray}
l_j=1-\frac{3\sigma_0}{4(\sigma_0+1)}(1-l_b)  \label{Psi_analyt},\\
\bar{\Gamma}=1+\frac{\sigma_0}{4}(1-l_b) \label{Gamm_analyt}.
\end{eqnarray}
Substituting Equation (\ref{Gamm_analyt}) into Equation (\ref{deta}) we obtain the Poynting flux profile:
\begin{equation}
-\left(\frac{\sigma_0}{4}+1\right)^2\ln l_b -\frac{\sigma_0}{8}(\sigma_0+4)(1-l_b)+\frac{\sigma_0^2}{32}(1-l_b^2)=2(x-1)
\label{L_B_analyt}
\end{equation}
at $x<x_{bc}$ and $l=l_{bc}$ at $x>x_{bc}$, where $l_{bc}=l_b(x=x_{bc})$ denotes the fraction of the Poynting flux that cannot dissipate. It is related 
to the critical magnetization $\sigma_c$ through $l_{bc}=(1+4/\sigma_0)/(1+4/\sigma_c)$.  It is readily seen that in general the asymptotic 
power, Lorentz factor and magnetization assume the bounds
\begin{eqnarray}
l_{j\infty}\ge1-\frac{3\sigma_0}{4(\sigma_0+1)}(1-l_{bc})  \label{Psi_infty},\\
\Gamma_\infty/\Gamma_0\ge1+\frac{\sigma_0}{4}(1-l_{bc}) \label{Gamm_infty},\\
\sigma_\infty=\sigma_0l_{bc}/\bar{\Gamma}_\infty\le\frac{4\sigma_0l_{bc}}{4+\sigma_0(1-l_{bc})}.\label{sig_infty}
\end{eqnarray}
In the case of complete dissipation, $l_{bc}=0$, the latter reduce to
\begin{eqnarray}
l_{j\infty}\ge\frac{\sigma_0+4}{4(\sigma_0+1)}  \label{Psi2_infty},\\
\Gamma_\infty\ge\Gamma_0(1+\sigma_0/4) \label{Gamm2_infty}.
\end{eqnarray}
Evidently, the asymptotic power lies in the range $1\ge l_{j\infty}\ge0.25$ for $0\le\sigma_0\le\infty$.  The radiative efficiency is limited
to $\epsilon_c=3\sigma_0/4(\sigma_0+1)\rightarrow 0.75$ at $\sigma_0\rightarrow\infty$.  This limit
can be approached in highly magnetized  flows provided  $\alpha f_s(x)$ remains large on scales over which complete dissipation of the
magnetic field occurs.   It is worth noting that the above analysis ignores Compton scattering off cold electrons in the flow, which is negligible
in the systems under consideration.

\begin{figure*}
\centerline{\includegraphics[scale=0.5]{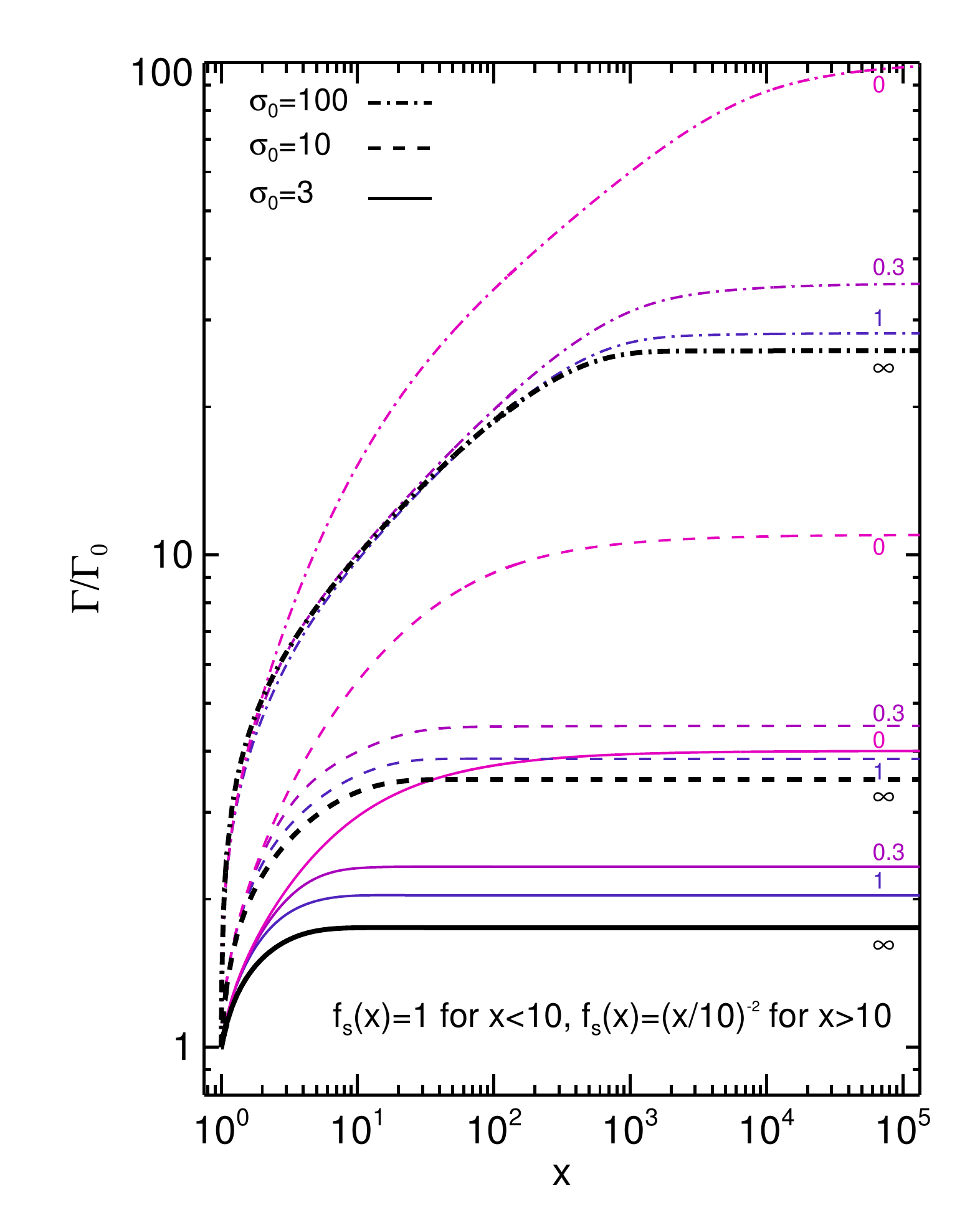}\includegraphics[scale=0.5]{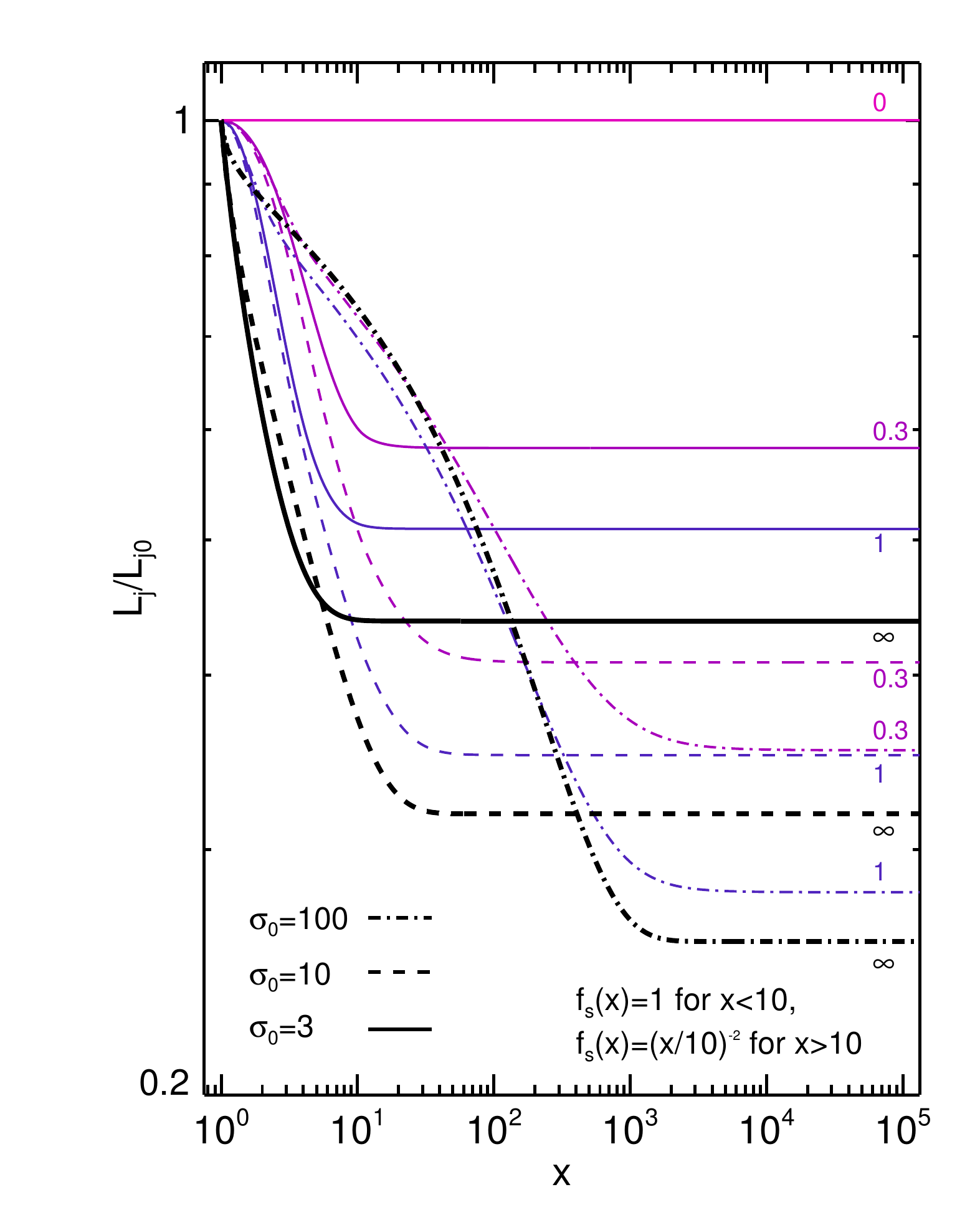}}
\caption{Same as figure \ref{fig-f1}, but for ambient radiation intensity profile $f_s(x)=1$ for $x<10$ and $f_s(x)=x^{-2}$ for $x\ge10$. }
\label{fig-f2}
\end{figure*}

Numerical solutions of Equations (\ref{dPsi})-(\ref{deta}) are exhibited in figures \ref{fig-f1} and \ref{fig-f2}, for models 1 and 2 respectively . 
The right panels delineate the evolution of the normalized total power $l_j$, and the left panels the evolution of the Lorentz factor 
$\Gamma/\Gamma_0$. 
Each group of lines of a given type correspond to solutions with the same initial magnetization $\sigma_0$ and different values of $\alpha$, 
as indicated.   The thick black lines correspond to the analytic solution given by Equations (\ref{Psi_analyt})-(\ref{L_B_analyt}). 
As seen, this solution is a good approximation in the regime $\alpha f_s(x)>1$.    We find that even for $\alpha<1$ substantial losses are expected.  For
example, for $\alpha=0.3$ in figure \ref{fig-f2} the radiative losses exceed $50\%$ for flows with $\sigma_0>>1$.

In cases where radiative losses are dominated by synchrotron and SSC emission, 
for which $\xi_c=0$, we recover the result
\begin{eqnarray}
&&l_{j}=1-\frac{\sigma_0}{2(\sigma_0+1)}(1-l_{b})  \label{Psi_no_fric},\\
&&\bar{\Gamma}=1+\frac{\sigma_0}{2}(1-l_{b}) \label{Gamm_no_fric},
\end{eqnarray}
obtained in DS02 in the case of strong radiative losses.   For complete dissipation, $l_{bc}=0$, the asymptotic values approach 
$l_{j\infty}\rightarrow 0.5$ and $\Gamma\rightarrow\Gamma_0(1+\sigma_0/2)$
in the limit $\sigma_0>>1$.

\section{Discussion}

Dissipative Poynting-flux  dominated flows can be very efficient emitters of electromagnetic radiation.  In sources whereby the
plasma in the reconnection zone cools predominantly through inverse Compton scattering of ambient radiation,
up to $75\%$ of the initial outflow power can be converted to gamma rays, provided that nearly complete dissipation of the
magnetic field occurs,  and that in the dissipation region the cooling rate exceeds the local dissipation rate.   If the cooling 
is dominated by synchrotron and SSC emission, the radiative efficiency is lower, $\epsilon_c \le 0.5$.

If dissipation commences at $\Gamma_0\sim$ a few, then we anticipate $\delta_B\simeq10^2-10^3 M$ for typical reconnection speeds
observed in recent numerical simulations, $v_r\simlt0.1 c$.   If the luminosity of the radiation intercepted by the jet is a fraction $\eta$ of
the Eddington value, then we estimate $\alpha\simeq 3\times10^{10}\eta (r_g/\delta_B)^2\simgt 10^4\eta$, 
independent of the mass of the central engine.   Thus, we anticipate radiative friction to be important 
in luminous, Galactic and extragalactic sources.   Detailed calculations (Joshi et al 2014) 
show that in a prototypical blazar,
like 3C279, the energy density of radiation intercepted by the jet is roughly constant, $u_s\simeq 10^{-3}$ ergs cm$^{-3}$, inside the broad line region, up
to a radius of $r\simeq10^{18}$ cm, and then declines roughly as $r^{-2}$.   This profile corresponds to our model 2
shown in figure \ref{fig-f2}.  Taking $\xi_e=0.5$ and a power law index $q=1.5$ for which $\chi=0.3$, we estimate $\alpha\simgt4$ from Equation ({\ref{alpha}).   From figure \ref{fig-f2} we expect high radiative efficiency in those 
objects, as indeed inferred from observations.  In the TeV blazars, synchrotron and SSC
emission most likely dominate.  The radiative efficiency  is then lower, but can still approach $25\%$ even if 
only half of the Poynting energy dissipates in the TeV emission zone.

An interesting possibilty is that complete magnetic field dissipation occurs, under certain conditions, 
in two distinct stages.   On small scales, the unstable magnetic field configuration established during the injection of the outflow
relaxes to a more stable configuration.  During this stage  gamma-ray emission is produced with high efficiency. 
Nontheless, the jet remains magnetically dominated if only a fraction of the Poynting flux can dissipate. 
On vastly larger scales, the outflow encounters a flat density profile medium and re-collimates.  If its
magnetization remains sufficiently high, $\sigma\sim$ a few, when reaching those scales, then its
subsequent evolution would depend on its relative power, as shown recently in the case
of AGN (Tchekhovskoy \& Bromberg 2015).  Powerful jets will not be affected significantly by the 
external medium and will propagate stably to large distances at relativistic speeds, 
forming strong shocks at the jet head.  At large viewing angles those appear as FRII radio sources.  Less powerful jets
are susceptible to the kink instability (Bromberg \& Tchekhovskoy 2015), leading to further dissipation
of the magnetic energy and wobbling of the jet head that slows it down. Those would appear as FRI radio sources.
In both cases we expect strongly beamed gamma-ray emission on sub-parsec and parsec scales, that can be detected in 
sources observed at small viewing angles, consistent the unified scheme for radio loud AGNs.
For instance,  if the initial magnetization is $\sigma_0=50$ and only half of the initial Poynting energy can 
dissipate during the first stage, then from Equation (\ref{Psi_infty})-(\ref{sig_infty}) 
we obtain gamma-ray production efficiency of $1-l_j\simeq0.37$, Lorentz $\Gamma\simeq 7\,\Gamma_0$ and magnetization
$\sigma\simeq 3.5$ at the end of the first stage.   If radiation drag is insignificant, as might be the case 
in the fainter sources, the gamma-ray production efficiency may be somewhat smaller, $0.25$, as
seen from Equation (\ref{Psi_no_fric}), and the Lorentz factor somewhat higher,
$\Gamma\simeq 12\,\Gamma_0$.   If the flow remains roughly conical during its subsequent evolution, then it will be 
magnetically dominated when encountering the flat density medium, and the analysis of  Tchekhovskoy \& Bromberg (2015),
as described above,  applies.

\section*{Acknowledgments}

We thank Lorenzo Sironi for useful comments.  \\
AL acknowledges the support of The Israel Science Foundation (grant 1277/13). NG acknowledges the support of the I-CORE Program of the Planning and Budgeting Committee, The Israel Science Foundation (grant 1829/12) and the Israel Space Agency (grant 3-10417).

\end{document}